\documentclass[%
 reprint,
superscriptaddress,
longbibliography,
 amsmath,amssymb,
 aps,
prb,
]{revtex4-1}
\setcitestyle{numbers,square}

\usepackage{braket}
\usepackage{graphicx}% Include figure files
\usepackage{dcolumn}% Align table columns on decimal point
\usepackage{bm}% bold math
\usepackage[normalem]{ulem} % strikethrough
\usepackage{color}

\usepackage{hyperref}% add hypertext capabilities
\hypersetup{
    colorlinks,%
    citecolor=blue,%
    linkcolor=blue,%
    urlcolor=blue
}

\newcommand{\alloy}{Pt(Hg$_{x}$Se$_{1-x}$)$_2$}

\newcommand{\orcid}[1]{\href{https://orcid.org/#1}{\includegraphics[width=8pt]{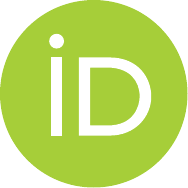}}}
\begin{document}

\title{Topological insulating phase arising in transition metal dichalcogenide alloy}

\author{F. Crasto de Lima\orcid{0000-0002-2937-2620}} 
\email{felipe.lima@ilum.cnpem.br}
\affiliation{Ilum School of Science, CNPEM, Campinas, SP, Brazil}

\author{B. Focassio\orcid{0000-0003-4811-7729}} 
\affiliation{Ilum School of Science, CNPEM, Campinas, SP, Brazil}
\affiliation{Center for Natural and Human Sciences, Federal University of ABC, Santo Andre, SP, Brazil}

\author{R. H. Miwa\orcid{0000-0002-1237-1525}}
\email{hiroki@ufu.br}
\affiliation{Instituto de F\'isica, Universidade Federal de Uberl\^andia, Uberl\^andia, MG, Brazil}

\author{A. Fazzio\orcid{0000-0001-5384-7676}}
\email{adalberto.fazzio@ilum.cnpem.br}
\affiliation{Ilum School of Science, CNPEM, Campinas, SP, Brazil}
\affiliation{Center for Natural and Human Sciences, Federal University of ABC, Santo Andre, SP, Brazil}

\date{\today}

\begin{abstract}

Transition metal dichalcogenides have been the subject of numerous studies addressing technological applications and fundamental issues. Single-layer PtSe$_2$ is a semiconductor with a trivial bandgap, in contrast, its counterpart with $25\%$ of Se atoms substituted by Hg, Pt$_2$HgSe$_3$ (jacutingaite, a naturally occurring mineral), is a 2D topological insulator with a large bandgap. Based on {\it ab-initio} calculations, we investigate the energetic stability, and the topological transition in Pt(Hg$_x$Se$_{1-x}$)$_2$ as a function of alloy concentration, and the distribution of Hg atoms embedded in the PtSe$_2$ host. Our findings reveal the dependence of the topological phase with respect to the alloy concentration and robustness with respect distribution of Hg. Through a combination of our {\it ab-initio} results and a defect wave function percolation model, we estimate the random alloy concentration threshold for the topological transition to be only $9\%$. Our results expand the possible search for non-trivial topological phases in random alloy systems.

\end{abstract}

\maketitle

\section{Introduction}

Finding new candidates for topological materials has become a critical point in the applications and technology advances \cite{NATmanipatruni2019}, given the need for (i) large topological gaps\,--\, dictating the electronic picture robustness; (ii) structural stability under atmospheric conditions; and (iii) accessible synthesis routes for scaling the production. The search for (new) materials fulfilling all those characteristics is an active area. Recently, a naturally occurring Pt-based mineral, Pt$_2$HgSe$_3$, was discovered, and later was characterized as hosting a large bulk bandgap 2D topological insulating phase [(i)] \cite{TNcabral2008, PRLmarrazo2018, PRBdelima2020}. This  natural occurrence suggests that Pt$_2$HgSe$_3$ (also known as jacutingaite) is structurally stable under atmospheric conditions [(ii)]. Its structure is made up of a transition  metal dichalcogenide (TMD) PtSe$_2$ backbone with 25\,\% of the chalcogen atoms substituted by Hg \cite{PRBdelima2020}. This implies that similar routes yielding PtSe2, or post-processing involving the incorporation of Hg onto the PtSe$_2$ matrix, can be used to obtain a large-scale production [(iii)] of such a 2D topological material \cite{PRLkomsa2012, PNASyunfan2021, AMqin2022}. Furthermore, the experimental knowledge on the synthesis, modification, and engineering of TMD-based devices makes this class of materials an obvious candidate for the transition from fundamental research to applications. For example, the synthesis of a fully metalized source/drain device based on 2D van der Waals  heterostructures of PtSe$_2$ \cite{ACSAMIdas2021}. 

Defects in 2D TMDs provide a controllable degree of freedom for tuning their electronic and structural properties \cite{2DMlin2016}. The incorporation of Hg atoms in the PtSe$_2$ matrix [$\rightarrow$ Pt(Hg$_x$Se$_{1-x}$)$_2$] can result in localized defects and disorder. Disorder effects may play a deleterious role in the topological phase in Bi$_2$Se$_3$ \cite{PRBfocassio2021}. On the other hand, it can regulate the transport mechanism \cite{NATCOMMqiu2013} or induce topological phases in 2D TMDs \cite{NLcrasto2021}. Indeed, the knowledge of the alloying and disordered effects at the atomic scale is a critical issue for predicting and synthesizing topological materials based on 2D TMDs.

In this paper, we have studied the quantum spin-Hall (QSH) phase in PtSe$_2$ monolayer (ML) upon the inclusion of substitutional Hg$_\text{Se}$ impurities, resulting in ordered and random Pt(Hg$_x$Se$_{1-x}$)$_2$ alloys. We start our investigation through the calculation of the Hg$_\text{Se}$ substitutional energy in the PtSe$_2$ host. We have discussed the energetic preference for the ordered $\rightarrow$ random transition as a function of the temperature and alloy concentration. Topological characterization of the ordered phase, performed by the calculation of the $\mathcal{Z}_2$ invariant, reveals that the  Pt(Hg$_x$Se$_{1-x}$)$_2$ alloys are topological as predicted by the Kane-Mele model, for $x$ between 0.25 and 0.75. Whereas, at the edge concentrations, we found the $\mathcal{Z}_2$-metallic phase \cite{PRBzhao2014} for $x$\,=\,1.0 (PtHg$_2$), and a trivial insulator for $x$\,=\,0, PtSe$_2$ host. The topological character of the random alloys, examined through the calculation of the spin-Bott index, reveals that the topological phase of the Pt(Hg$_x$Se$_{1-x}$)$_2$ alloys has been preserved even upon a random distribution of the Hg$_\text{Se}$ impurities. In the sequence, the topological non-trivial $\leftrightarrow$ trivial transition has been examined by the calculation of the percolation threshold of the orbital overlap of the Hg$_\text{Se}$ wave functions, randomly, embedded in the PtSe$_2$ host.

\section{Methods}

The calculations were performed on the Viena ab-initio simulation package (VASP) \cite{vasp1}, within a plane-wave base with a cutoff energy of $400$\,eV. The spin-orbit coupling was taken into account in all calculations, where the exchange and correlations were treated with the Perdew-Burk-Ernzehof functional \cite{PBE}. The electron-ion interactions were described within the projected augmented wave (PAW) \cite{PRBblochl1994} method, with all atoms allowed to relax until each atom resultant force was lower than $10^{-2}$\,eV/{\AA}. For the pristine unity cell, the total energy was calculated on the 2D Brillouin zone with a regular K-point grid of $5 \times 5$ and $7 \times 7$ for the atom's relaxation and self-consistent charge density, respectively. The same K-point density was taken on the larger supercell calculations. The topological invariant ($\mathcal{Z}_2$ \cite{PRBsoluyanov2011, PRByu2011} and spin-Bott index \cite{PRBhuang2018}) and semi-infinite ribbon calculations were performed after extracting a Wannier function tight-biding model from the ab-initio calculations through the Wannier90 code \cite{CPCmostofi2014}.

For the random alloy, we have generated the alloy structure through the Special-quasirandom Structure (SQS) \cite{PRLzunger1990} for cells with 48 and 27 atoms achieving 1st-neighbor pair correlation (PC) exactly random, and 2nd-neighbor and 3rd-neighbor PC exactly random for most structures. Such structures represent systems without translational symmetry, their reciprocal space is undefined, and therefore a different approach for computing the $\mathcal{Z}_2$ invariant is needed. Recently an approach involving the evaluation of the spin Bott index was successful to characterize non-periodic systems \cite{PRBhuang2018, PRBfocassio2021}, which we computed for the random alloy structures.

\section{Results}

\subsection{Hg-Hg-interaction embedded in PtSe$_2$ matrix}

The  formation of jacutingaite, Pt$_2$HgSe$_3$, can be viewed as a (partial) substitution of Se with Hg atoms in the 1T phase of the  Pt$_2$Se matrix \cite{TNcabral2008, PRLmarrazo2018, PRBdelima2020}. In this case, the substitutional energy ($E_s$)  of $n$ Se atoms by Hg in the PtSe$_2$ is
\begin{equation}
E_s= (E_{\rm system} + n \times \mu_{\rm Se} - E_{\rm matrix} - n \times \mu_{\rm Hg})/n, \label{eq-d}
\end{equation}
with $E_{\rm system}$ the total energy of the final (Hg$_\text{Se}$ doped) system, $E_{\rm matrix}$ the total energy of the 1T Pt$_2$Se matrix; $\mu_{\rm Se}$ and $\mu_{\rm Hg}$ are the chemical potentials of Se and Hg atoms, respectively. 

\begin{figure}[h!]
\includegraphics[width=\columnwidth]{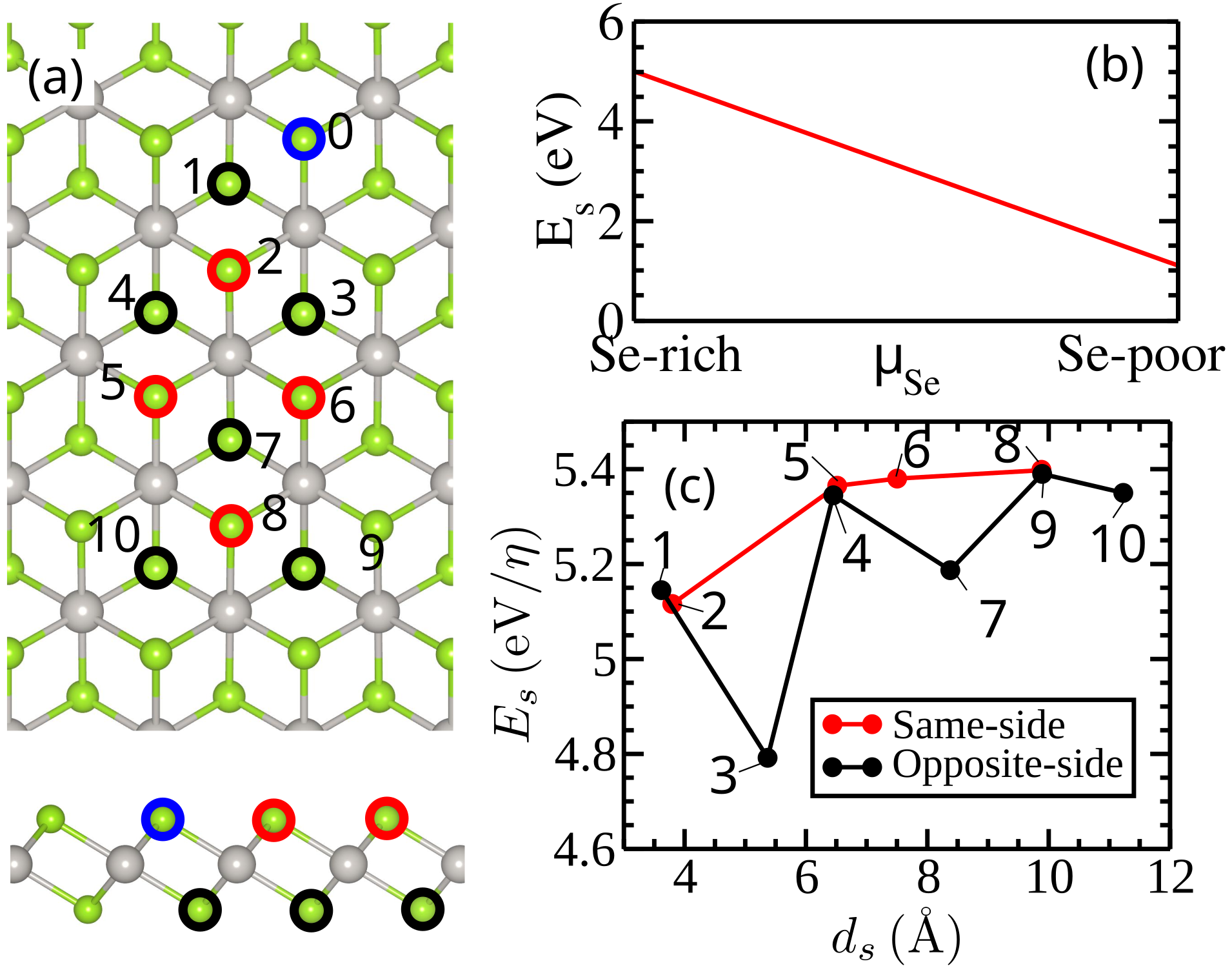}
\caption{\label{Hg-Hg-int} (a) Substitutional sites for Hg on PtSe$_2$ matrix. (b) Se substitutional energy for a single Hg (Eq.~\ref{eq-d}) as a function of Se chemical potential. (c) Substitutional energy for two Hg atoms separated by a distance $d_s$, the number labels the site of figure (a) with red/black curves of the pair of Hg lying on the same/opposite side of the PtSe$_2$ layer.}
\end{figure}

Let us start with a single substitutional impurity of Hg in the Se site, Hg$_\text{Se}$. The calculations were performed by using a Pt$_2$Se tetragonal supercell with dimensions of $18.7$\,{\AA} and $32.4$\,{\AA}, schematically shown in Fig.\,\ref{Hg-Hg-int}(a), in order to avoid  spurious interactions between the periodic images. Our results of $E_s$, presented in Fig.\,\ref{Hg-Hg-int}(b), indicates that the formation of a single Hg$_\text{Se}$ [indicated by a blue-circle in Fig.\,\ref{Hg-Hg-int}(a)] is an endothermic process, $E_s>0$, with higher occurrence rate at the Se-poor (Pt-rich) condition. Although the positive values of $E_s$, it is worth noting that the combination of the energetic preference of formation of Se vacancies (V$_\text{Se}$) in PtSe$_2$  at the Se-poor condition \cite{absor2017defect, freire2022vacancy}, and the experimental realization of doping of TMDs (including PtSe$_2$) by filling the chalcogen vacancies created  by electron-irradiation \cite{PRLkomsa2012, AMqin2022}, makes the formation of Hg$_\text{Se}$ in PtSe$_2$ a quite likely process.    

%%%%%%FIG1

In the sequence, the higher mobility of V$_\text{Se}$, induced by an electron beam, promotes the formation of energetically stable paired divacancies and vacancy-line structures in single layer PtSe$_2$ \cite{NATMATlin2017, NLchen2022}.  Since, as discussed above, these intrinsic defects can be filled up by foreign atoms, it is worth investigating the interaction between the Hg$_\text{Se}$ impurities embedded in  PtSe$_2$. Here, we have calculated the formation energy of Hg$^0_\text{Se}$-Hg$^i_\text{Se}$ pair configurations, with $i$\,=\,1-10 as depicted in Fig.\,\ref{Hg-Hg-int}(a).  Black (Red) circles indicate Hg$^i_\text{Se}$ atoms lying on the same (opposite) side of Hg$^0_\text{Se}$ (blue circle). Our results of $E_s$ as a function of the  distance between the impurities, Fig.\,\ref{Hg-Hg-int}(c), reveal that the  Hg$^0_\text{Se}$-Hg$^3_\text{Se}$ configuration will be the most stable (likely) one, which is characterized by in-line Hg-Pt-Hg bonds sharing the same Pt atoms. Interestingly, such a Hg-Pt-Hg bonding configuration is exactly that found in the Pt$_2$HgSe$_3$ crystals, which suggests that Hg$^0_\text{Se}$-Pt-Hg$^3_\text{Se}$ pairs act  as seeds for the formation of topological insulator jacutingaite crystals. On the other hand, in contrast with the  Pt$_2$HgSe$_3$ crystals, we may have (i) other concentrations,  as well as (ii) periodic or random distribution of Hg$_\text{Se}$ impurities embedded in PtSe$_2$. In the sequence, we will discuss the importance of (i) and (ii) to the electronic properties of Hg-doped PtSe$_2$ monolayers.

\subsection{Ordered and random alloys}

%%%%%%FIG2
\begin{figure}[h!]
\includegraphics[width=\columnwidth]{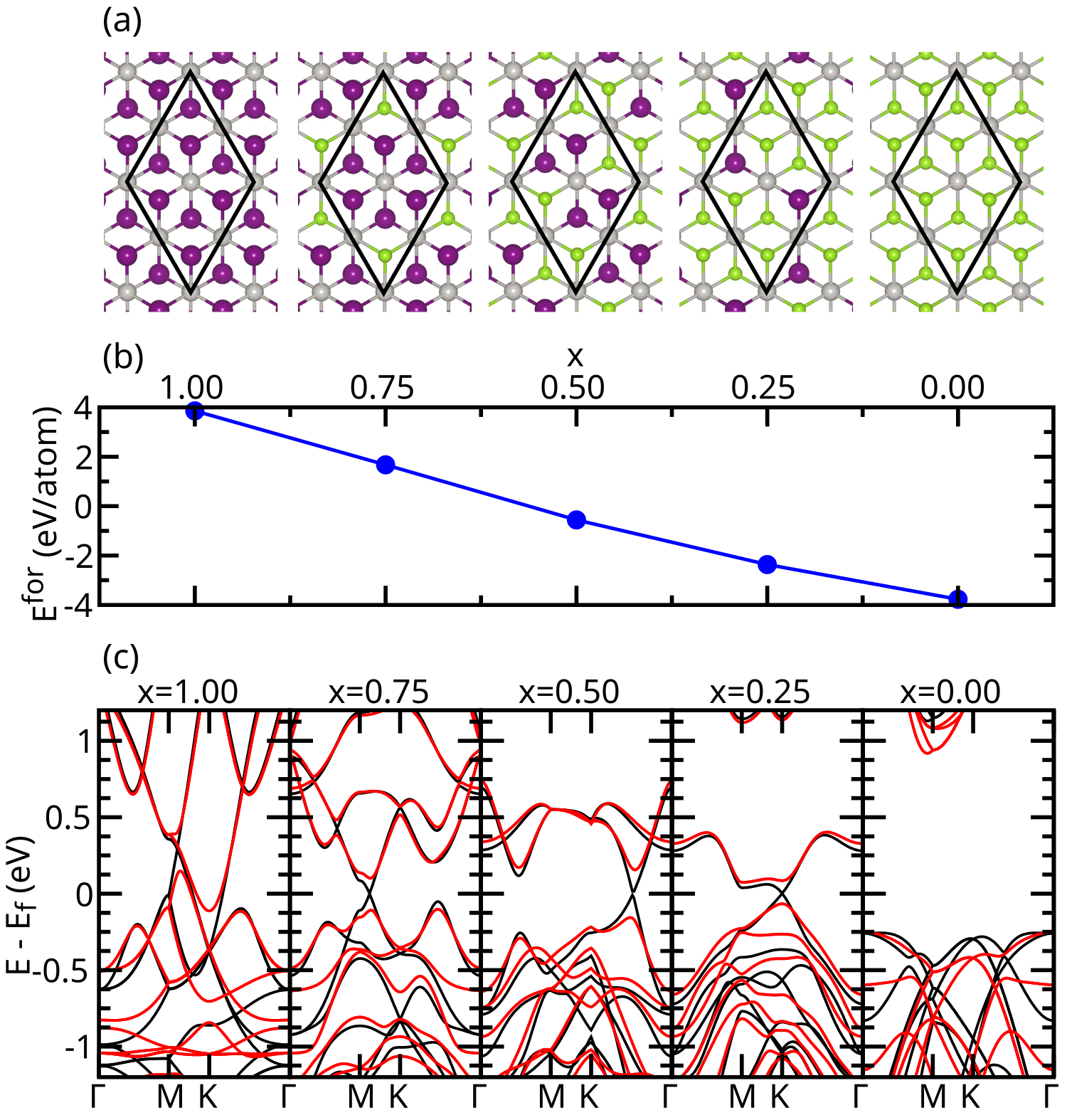}
\caption{\label{ordered} Ordered configuration (a) atomic structure, (b) formation energy, and (c) band structure for {\alloy}. The band shown in black (red) are for calculation without (with) SOC.}
\end{figure}

%%%%%%FIG3
\begin{figure*}[t]
\includegraphics[width=1.6\columnwidth]{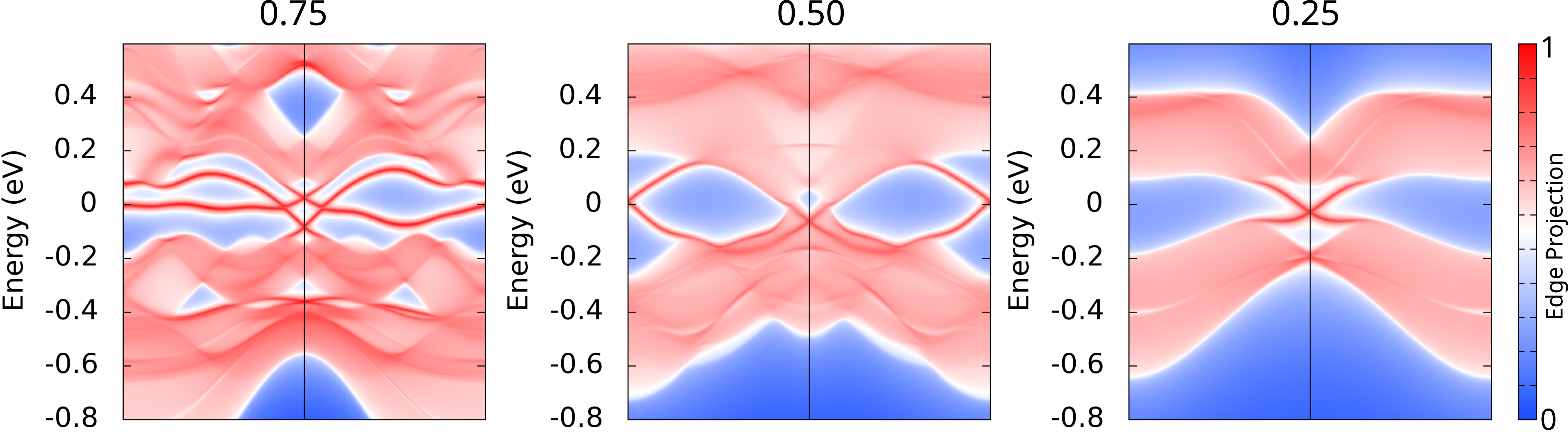}
\caption{\label{edge-states} Edge states for ordered structure. All of the explored systems, except the x=0 and x=1, are topologically non-trivial.}
\end{figure*}

%%%%%%FIG4
\begin{figure}[h!]
\includegraphics[width=0.9\columnwidth]{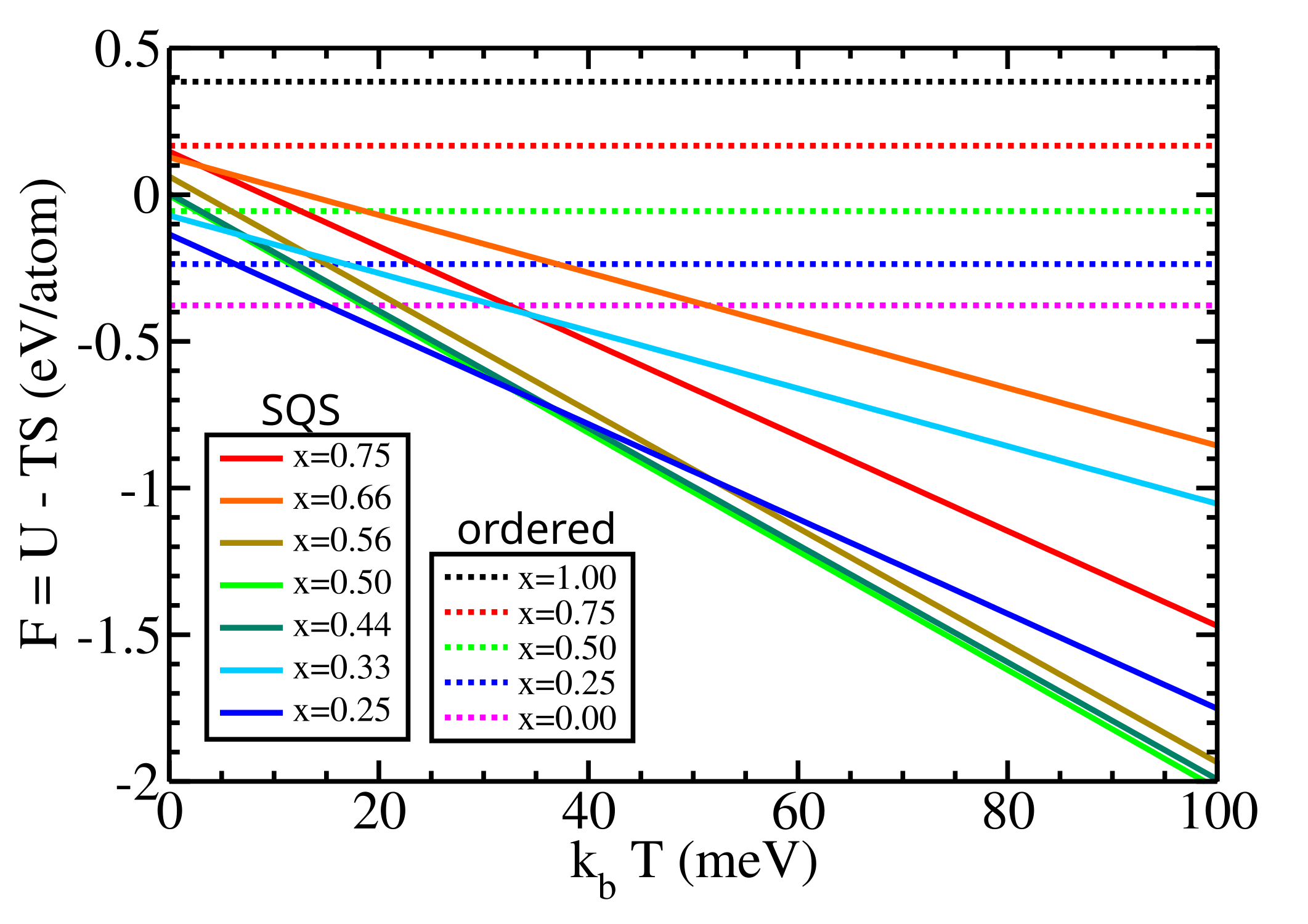}
\caption{\label{free-energy} Free energy as a function of temperature with configurational entropy accounted for the ordered (dotted lines) and SQS (continuous lines) alloys.}
\end{figure}

We start by examining ordered Pt(Hg$_x$Se$_{1-x}$)$_2$ alloys, with $x$ $=$ 0, 0.25, 0.50, 0.75, and 1.0 created by incorporating Hg$_\text{Se}^0$-Pt-Hg$_\text{Se}^3$ units [Fig.\,\ref{Hg-Hg-int}] in the PtSe$_2$ host, Fig.\,\ref{ordered}(a). Pristine PtSe$_2$ ($x$\,=\,0) is energetically stable, characterized by negative values of formation energy, $E^{\rm for}$, $-0.48$\,eV/atom, Fig.\,\ref{ordered}(b), followed by the jacutingaite phase ($x$\,=\,0.25), which also present $E^{\rm for}$\,$<$\,0. These results are compatible with the natural formation of both the PtSe$_2$ and Pt$_2$HgSe$_3$ systems. Whereas for $x$\,=\,0.5 we find that PtHgSe lies on the verge of positive values  of $E^{\rm for}$.

%%%%%%FIG5
\begin{figure*}[tb]
\includegraphics[width=2\columnwidth]{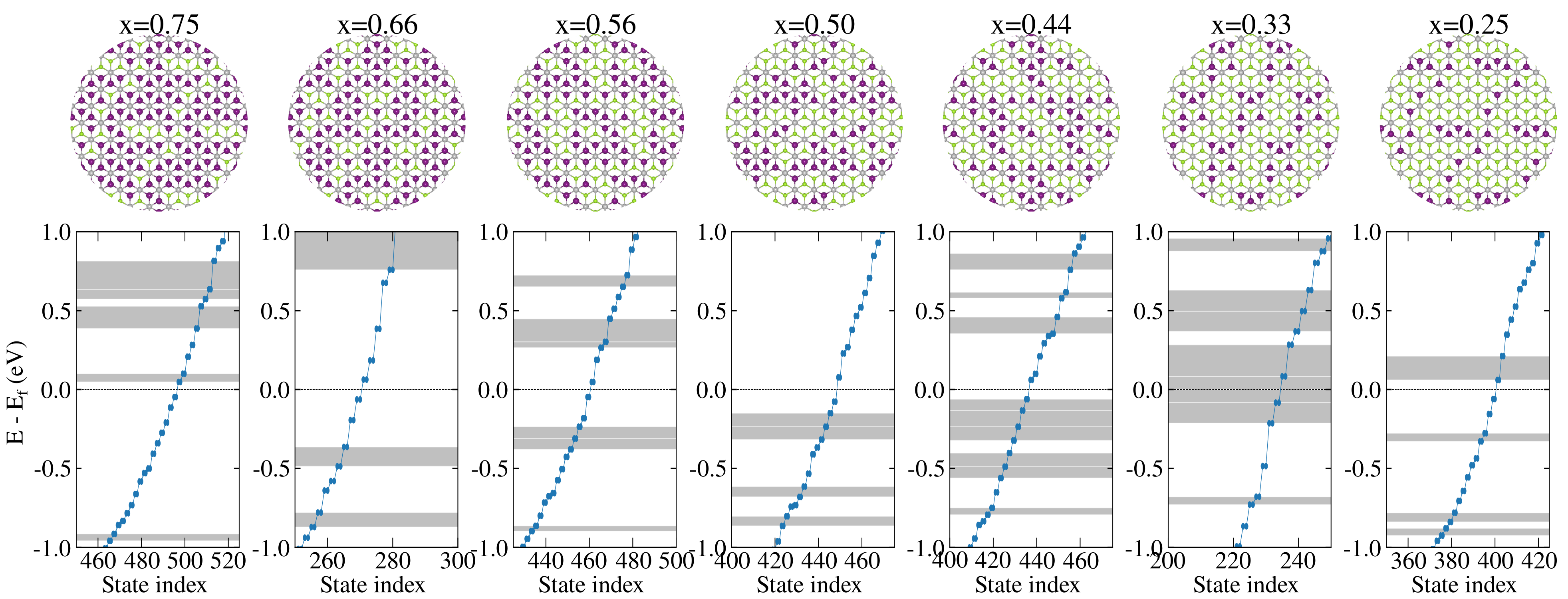}
\caption{\label{alloy-topo} Disordered SQS structures (upper panels) and respective topological gaps (lower panels).}
\end{figure*}

The electronic band structures of the ordered Pt(Hg$_x$Se$_{1-x}$)$_2$ alloys, presented in Fig.\,\ref{ordered}(c), show that the pristine  PtSe$_2$ host ($x$\,=\,0) is a semiconductor with a band gap of $1.2$\,eV, while at the other end ($x$\,=\,1) we find that PtHg$_2$ presents a metallic character. It is important to note that the SOC controls the semiconducting character in these Pt(Hg$_x$Se$_{1-x}$)$_2$ alloys, which were computed for $x$\,=\,0.25, 0.50, and 0.75. As seen by the solid black lines in Fig.\,\ref{ordered}(c), such alloys become semi-metallic by turning off the SOC, resulting in the linear crossing of energy bands at the Fermi level. Such a bandgap ($E_g$) opening, mediated by the SOC, is in consonance with the QSH effect as predicted by the Kane-Mele model. In particular, for jacutingaite, $x$\,=\,0.25, our DFT-SOC results of $E_g$ (=0.14\,eV)  are in good agreement with the one obtained by Marrazzo et al. \cite{PRLmarrazo2018}. Meanwhile, for the other concentrations, we found $E_g$ of 0.11 and  and 0.20\,eV for $x$\,=\,0.50, and 0.75, respectively. Interestingly, the SOC-induced bandgaps do not exhibit a nearly linear behavior with respect to the alloy concentration, in contrast to the results obtained for the formation energy, Fig.\,\ref{ordered}(b) and (c).

In order to provide further confirmation of the non-trivial topology of Pt(Hg$_x$Se$_{1-x}$)$_2$, we have calculated the $\mathcal{Z}_2$ invariant by extrapolating the system to a Wannier base representation and analyzing its spatial evolution through the Brillouin zone (BZ) \cite{PRByu2011}. Our results reveal that  the Pt(Hg$_x$Se$_{1-x}$)$_2$ ordered alloys present a non-trivial topological phase [except the pristine $x=0$ case]. Here, although the full Hg-covered system is metallic, we considered that the negative dispersive band at the M-point is fully occupied, while the positive dispersive band at K-point is fully empty, with an arising $\mathcal{Z}_2$-metallic topological phase \cite{PRBzhao2014}. Finally, as a signature of the SOC-induced topological gaps in Pt(Hg$_x$Se$_{1-x}$)$_2$,  in Fig.~\ref{edge-states} we present the projected edge states which are characterized by the  emergence of (chiral) spin-polarized Dirac cones characteristic of the QSH phase.

According to Refs. \cite{PRBfocassio2021, NATPHYSmitchell2018, PRLagarwala2017, NLcrasto2021, NLcosta2019}, disordered systems can also exhibit non-trivial topological phases. Here, focusing on the Pt(Hg$_x$Se$_{1-x}$)$_2$ alloys, we might have a random distribution of the Hg$_\text{Se}$ impurities preserving the host's (1T) backbone structure. Such randomness can alter the electronic/topological properties/identities of Pt(Hg$_x$Se$_{1-x}$)$_2$ alloys as well as the energetic stability for each alloy concentration. 

In order to provide a realistic description of random alloys, the distribution of the Hg$_\text{Se}$ impurities in Pt(Hg$_x$Se$_{1-x}$)$_2$ were generated by using the SQS approach \cite{PRLzunger1990}. We have generated SQS structures ranging from $x$\,=\,0.25 to $x$\,=\,0.75. By taking into account the configurational entropy, $S=k_B \ln \left(\Omega \right)$, with $\Omega$ the number of equivalent configurations, we have determined the free energy ($F=U-TS$) as a function of the temperature ($T$) and alloy concentration, $x$. Our results of free energy, summarized in Fig.\,\ref{free-energy}, reveal that (i) there is an energetic preference for the  pristine PtSe$_2$ ($x=0$) for temperatures with $k_B T<15$\,meV;  while (ii) the random counterpart of jacutingaite becomes more stable than its ordered phase for $k_B T>6$\,meV; where (iii) such an energetic preference for $x$\,=\,0.25 has been maintained up to temperatures with $k_B T<32$\,meV; and (iv) above that, we find the  PtHgSe random alloy will present the lowest free energy.

As discussed above, we found that  Pt(Hg$_{x}$Se$_{1-x}$)$_2$ ordered alloys, with $x$\,=\,0.25, 0.50, and 0.75, present the QSH phase characterized by the SOC-induced non-trivial bandgaps at the Fermi level, Fig.\,\ref{ordered}(c). In the sequence, we will examine the topological identities of Pt(Hg$_{x}$Se$_{1-x}$)$_2$ random alloys within the same range of alloy concentrations. Given the non-periodicity of the random alloys, we have computed a real space invariant (Spin Bott-index) which is equivalent to the spin-Chern number \cite{PRBhuang2018}. In Fig.~\ref{alloy-topo} we highlight (in gray) the topological energy gaps of the random alloy systems. Our results reveal that, although the random distribution introduces trivial bandgaps close to the Fermi energy, the topological gaps are always present within an energy interval of $\pm$\,0.5\,eV around the Fermi level. Here, although the non-uniform distribution of the Hg$_\text{Se}$ impurities gives rise to sensible differences in the energy localization  and the width of the topological gaps, our findings indicate that the  Pt(Hg$_{x}$Se$_{1-x}$)$_2$ system will always exhibit non-trivial bandgaps for alloy concentrations between 0.25 and 0.75, regardless of how  (ordered or randomly) the Hg$_\text{Se}$ impurities are distributed in the PtSe$_3$ host. This result is in line with that obtained by Lee\, et al. \cite{JPCMlee2014}; the authors verified the  stability of the $\mathcal{Z}_2$-topological phase upon the inclusion of aleatory vacancies, through disordered potentials, in the Kane-Mele model on the honeycomb lattice. Back to our Pt(Hg$_{x}$Se$_{1-x}$)$_2$ system, since PtSe$_2$ ($x$\,=\,0) is a trivial insulator, while Pt$_2$HgSe$_3$ ($x$\,=\,0.25, jacutingaite) is a topological insulator, it is expected a trivial $\leftrightarrow$ non-trivial topological transition for alloy concentrations between 0 and 0.25.

\subsection{Topological limit}

Lowering the Hg$_\text{Se}$ concentration, $x \rightarrow 0$, will lead the Pt(Hg$_{x}$Se$_{1-x}$)$_2$ system to a topologically trivial phase, PtSe$_2$. In this subsection, we will look at a trivial $\leftrightarrow$ non-trivial topological transition in light of the spatial localization of the Hg$_\text{Se}$ impurities (electron percolation limit).

The topological character can be predicted by the spin Chern number, $c^{(s)} =  c_+ - c_-$, with $c_j$ the Chern number on the subspace of $j=+,-$ spin.  This is dependent on the integration over a closed surface of the Berry connection,
\begin{equation}
    c_j = \sum_{n}^{\rm occup}\oint_C \vec{A}_n^{(j)} \cdot d\vec{k},
\end{equation}
with $\vec{A}_n^{(\sigma)} = i\braket{n,\vec{k},\sigma| \nabla_{k} | n, \vec{k},\sigma}$. Here, $i \nabla_k$ is the definition of the position operator on the momentum space. If the system is composed only of fully  localized states (atomic limit), that is, eigenfunctions of the position operator,
\begin{equation}
    c_j = \sum_{n}^{\rm occup} x_n \oint_C dk = 0.
\end{equation}
Therefore, a non-zero Chern number requires the electronic states to not be completely localized. This imposes a necessary (however not sufficient) condition for the emergence of the QSH phase. That is, in Pt(Hg$_{x}$Se$_{1-x}$)$_2$ alloys, the Hg$_\text{Se}$ impurity wave functions (IWFs) should interact with one another above a given percolation limit, i.e. a threshold random alloy concentration which leads to the trivial $\leftrightarrow$ non-trivial topological transition.

Here we can estimate such a threshold concentration in  Pt(Hg$_{x}$Se$_{1-x}$)$_2$ random alloys based upon the percolation threshold of the IWFs. In this case, the topological transition will be dictated by the electronic percolation through impurity sites embedded in the PtSe$_2$ host. That is if the overlap of neighboring IWFs enables the electronic percolation through a large (infinite) clustered set of Hg$_\text{Se}$ atoms, the energy dispersion between the impurity states will lead to the non-trivial topological phase. On the other hand, if the neighboring IWFs do not overlap  (atomic limit), the absence or nearly zero energy dispersion will result in a trivial phase. In this case, based on the random distribution of Hg$_\text{Se}$ in a hexagonal lattice (on the Se sites), the percolation threshold can be written as \cite{PREsuding1999}, 
\begin{equation}
    \phi_c = 1 - e^{-n_C} = 0.69,
\end{equation}
where $n_C$ is the critical ratio between the area ($A$) occupied by the IWFs and that of the unity cell (UC), $n_C=A_\text{I}/A_\text{UC}$. Here, the interaction between Hg$_\text{Se}$ impurities was estimated based on the substitutional energy, $E_s$, and electronic interaction between the impurity atoms. As shown in Fig.\,\ref{Hg-Hg-int}(c), the Hg$^0_\text{Se}$-Hg$^8_\text{Se}$, Hg$^0_\text{Se}$-Hg$^9_\text{Se}$, and  Hg$^0_\text{Se}$-Hg$^{10}_\text{Se}$ configurations have nearly the same $E_s$ values, i.e. substitutional energy differences ($\Delta E_s$) of about $0.05$\,eV/Hg$_\text{Se}$-atom, which corresponds to Hg$_\text{Se}$-Hg$_\text{Se}$ distances, $d_s$, between $10$ and $11$\,\AA. For $d_s$ of  $11$\,\AA, we found an electronic (HOMO-LUMO) interaction of $\sim 5$\,meV between the  Hg$_\text{Se}$ impurities, $\Delta\varepsilon=5$\,meV, which is smaller than the SOC induced non-trivial bandgaps. Thus, we can assume that $11$\,\AA\, is the lower limit for the orbital localization of the IWFs, i.e. for $d_s>11$\,\AA\ we will have  $\Delta E_s<0.05$\,eV/Hg$_\text{Se}$-atom and $\Delta\varepsilon<5$\,meV. In  Pt(Hg$_{x}$Se$_{1-x}$)$_2$, an averaged distance of $11$\,\AA\ between Hg$_\text{Se}$-Hg$_\text{Se}$ impurities results in a trivial $\leftrightarrow$ topological threshold concentration $x_t=0.1$. 

Here, our simulations going down to $x=0.25$ for the random alloy confirm the persistence of the topological phases within the threshold concentration. It is worth pointing out that simulating realistic disordered systems, within the SQS approach for concentrations below $x_{t}=0.09$, increases the computational costs considerably. That is, to achieve concentrations near the pristine case, the supercell size needs to be increased to achieve a good correlation towards real disordered systems. However, note that the Jacutingaite phase (ordered $x=0.25$) is the Kane-Mele model's realization. It is shown that randomly vacancies on the honeycomb Kane-Mele model (replacing Hg by Se) a possible topological phase is preserved down to $x \sim 0.2$ \cite{JPCMlee2014}. 

\section{Conclusion}

Based on {\it ab-initio} calculations, we have studied the energetic stability, and the trivial $\rightarrow$ non-trivial topological phase transition in single layer PtSe$_2$ mediated by substitutional Hg$_\text{Se}$ atoms, Pt(Hg$_x$Se$_{1-x}$)$_2$ alloys. We found an energetic preference for a random distribution of Hg$_\text{Se}$ with $x=25\%$  (jacutingaite's stoichiometry), with respect to the PtSe$_2$ host, for temperatures with $k_BT>15$\,meV. Meanwhile, random alloys with $x=50\%$, PtHgSe, become  more stable for $k_BT>32$\,meV. The robustness of the QSH phase against the random distribution of Hg$_\text{Se}$ substitutional atoms has been verified for alloy concentrations between $25\%$ and $75\%$. In addition, based on a combination of {\it ab-initio} results and a percolation model, we estimate a threshold concentration, of about 10\%, for the  topological non-trivial $\leftrightarrow$ trivial transition in Pt(Hg$_x$Se$_{1-x}$)$_2$ random alloys.

\begin{acknowledgments}

The authors acknowledge financial support from the Brazilian agencies FAPESP (grants 19/20857-0, 19/04527-0, and 17/02317-2), FAPEMIG, INCT-Nanocarbono, and Laborat\'{o}rio Nacional de Computa\c{c}\~{a}o Cient\'{i}fica for computer time (project ScafMat2).

\end{acknowledgments}

%\appendix

\bibliography{bib}% Produces the bibliography via BibTeX.

\begin{thebibliography}{30}%
\makeatletter
\providecommand \@ifxundefined [1]{%
 \@ifx{#1\undefined}
}%
\providecommand \@ifnum [1]{%
 \ifnum #1\expandafter \@firstoftwo
 \else \expandafter \@secondoftwo
 \fi
}%
\providecommand \@ifx [1]{%
 \ifx #1\expandafter \@firstoftwo
 \else \expandafter \@secondoftwo
 \fi
}%
\providecommand \natexlab [1]{#1}%
\providecommand \enquote  [1]{``#1''}%
\providecommand \bibnamefont  [1]{#1}%
\providecommand \bibfnamefont [1]{#1}%
\providecommand \citenamefont [1]{#1}%
\providecommand \href@noop [0]{\@secondoftwo}%
\providecommand \href [0]{\begingroup \@sanitize@url \@href}%
\providecommand \@href[1]{\@@startlink{#1}\@@href}%
\providecommand \@@href[1]{\endgroup#1\@@endlink}%
\providecommand \@sanitize@url [0]{\catcode `\\12\catcode `\$12\catcode
  `\&12\catcode `\#12\catcode `\^12\catcode `\_12\catcode `\%12\relax}%
\providecommand \@@startlink[1]{}%
\providecommand \@@endlink[0]{}%
\providecommand \url  [0]{\begingroup\@sanitize@url \@url }%
\providecommand \@url [1]{\endgroup\@href {#1}{\urlprefix }}%
\providecommand \urlprefix  [0]{URL }%
\providecommand \Eprint [0]{\href }%
\providecommand \doibase [0]{http://dx.doi.org/}%
\providecommand \selectlanguage [0]{\@gobble}%
\providecommand \bibinfo  [0]{\@secondoftwo}%
\providecommand \bibfield  [0]{\@secondoftwo}%
\providecommand \translation [1]{[#1]}%
\providecommand \BibitemOpen [0]{}%
\providecommand \bibitemStop [0]{}%
\providecommand \bibitemNoStop [0]{.\EOS\space}%
\providecommand \EOS [0]{\spacefactor3000\relax}%
\providecommand \BibitemShut  [1]{\csname bibitem#1\endcsname}%
\let\auto@bib@innerbib\@empty
%</preamble>
\bibitem [{\citenamefont {Manipatruni}\ \emph {et~al.}(2019)\citenamefont
  {Manipatruni}, \citenamefont {Nikonov}, \citenamefont {Lin}, \citenamefont
  {Gosavi}, \citenamefont {Liu}, \citenamefont {Prasad}, \citenamefont {Huang},
  \citenamefont {Bonturim}, \citenamefont {Ramesh},\ and\ \citenamefont
  {Young}}]{NATmanipatruni2019}%
  \BibitemOpen
  \bibfield  {author} {\bibinfo {author} {\bibfnamefont {Sasikanth}\
  \bibnamefont {Manipatruni}}, \bibinfo {author} {\bibfnamefont {Dmitri~E.}\
  \bibnamefont {Nikonov}}, \bibinfo {author} {\bibfnamefont {Chia-Ching}\
  \bibnamefont {Lin}}, \bibinfo {author} {\bibfnamefont {Tanay~A.}\
  \bibnamefont {Gosavi}}, \bibinfo {author} {\bibfnamefont {Huichu}\
  \bibnamefont {Liu}}, \bibinfo {author} {\bibfnamefont {Bhagwati}\
  \bibnamefont {Prasad}}, \bibinfo {author} {\bibfnamefont {Yen-Lin}\
  \bibnamefont {Huang}}, \bibinfo {author} {\bibfnamefont {Everton}\
  \bibnamefont {Bonturim}}, \bibinfo {author} {\bibfnamefont {Ramamoorthy}\
  \bibnamefont {Ramesh}}, \ and\ \bibinfo {author} {\bibfnamefont {Ian~A.}\
  \bibnamefont {Young}},\ }\bibfield  {title} {\enquote {\bibinfo {title}
  {{Scalable energy-efficient magnetoelectric spin--orbit logic}},}\ }\href
  {\doibase 10.1038/s41586-018-0770-2} {\bibfield  {journal} {\bibinfo
  {journal} {Nature}\ }\textbf {\bibinfo {volume} {565}},\ \bibinfo {pages}
  {35--42} (\bibinfo {year} {2019})}\BibitemShut {NoStop}%
\bibitem [{\citenamefont {Cabral}\ \emph {et~al.}(2008)\citenamefont {Cabral},
  \citenamefont {Galbiatti}, \citenamefont {Kwitko-Ribeiro},\ and\
  \citenamefont {Lehmann}}]{TNcabral2008}%
  \BibitemOpen
  \bibfield  {author} {\bibinfo {author} {\bibfnamefont {A.~R.}\ \bibnamefont
  {Cabral}}, \bibinfo {author} {\bibfnamefont {H.~F.}\ \bibnamefont
  {Galbiatti}}, \bibinfo {author} {\bibfnamefont {R.}~\bibnamefont
  {Kwitko-Ribeiro}}, \ and\ \bibinfo {author} {\bibfnamefont {B.}~\bibnamefont
  {Lehmann}},\ }\bibfield  {title} {\enquote {\bibinfo {title} {{Platinum
  enrichment at low temperatures and related microstructures, with examples of
  hongshiite (PtCu) and empirical 'Pt2HgSe3' from Itabira, Minas Gerais,
  Brazil}},}\ }\href {\doibase
  https://doi.org/10.1111/j.1365-3121.2007.00783.x} {\bibfield  {journal}
  {\bibinfo  {journal} {Terra Nova}\ }\textbf {\bibinfo {volume} {20}},\
  \bibinfo {pages} {32--37} (\bibinfo {year} {2008})}\BibitemShut {NoStop}%
\bibitem [{\citenamefont {Marrazzo}\ \emph {et~al.}(2018)\citenamefont
  {Marrazzo}, \citenamefont {Gibertini}, \citenamefont {Campi}, \citenamefont
  {Mounet},\ and\ \citenamefont {Marzari}}]{PRLmarrazo2018}%
  \BibitemOpen
  \bibfield  {author} {\bibinfo {author} {\bibfnamefont {Antimo}\ \bibnamefont
  {Marrazzo}}, \bibinfo {author} {\bibfnamefont {Marco}\ \bibnamefont
  {Gibertini}}, \bibinfo {author} {\bibfnamefont {Davide}\ \bibnamefont
  {Campi}}, \bibinfo {author} {\bibfnamefont {Nicolas}\ \bibnamefont {Mounet}},
  \ and\ \bibinfo {author} {\bibfnamefont {Nicola}\ \bibnamefont {Marzari}},\
  }\bibfield  {title} {\enquote {\bibinfo {title} {{Prediction of a Large-Gap
  and Switchable Kane-Mele Quantum Spin Hall Insulator}},}\ }\href {\doibase
  10.1103/PhysRevLett.120.117701} {\bibfield  {journal} {\bibinfo  {journal}
  {Phys. Rev. Lett.}\ }\textbf {\bibinfo {volume} {120}},\ \bibinfo {pages}
  {117701} (\bibinfo {year} {2018})}\BibitemShut {NoStop}%
\bibitem [{\citenamefont {de~Lima}\ \emph {et~al.}(2020)\citenamefont
  {de~Lima}, \citenamefont {Miwa},\ and\ \citenamefont
  {Fazzio}}]{PRBdelima2020}%
  \BibitemOpen
  \bibfield  {author} {\bibinfo {author} {\bibfnamefont {F.~Crasto}\
  \bibnamefont {de~Lima}}, \bibinfo {author} {\bibfnamefont {R.~H.}\
  \bibnamefont {Miwa}}, \ and\ \bibinfo {author} {\bibfnamefont
  {A.}~\bibnamefont {Fazzio}},\ }\bibfield  {title} {\enquote {\bibinfo {title}
  {{Jacutingaite-family: A class of topological materials}},}\ }\href {\doibase
  10.1103/PhysRevB.102.235153} {\bibfield  {journal} {\bibinfo  {journal}
  {Phys. Rev. B}\ }\textbf {\bibinfo {volume} {102}},\ \bibinfo {pages}
  {235153} (\bibinfo {year} {2020})}\BibitemShut {NoStop}%
\bibitem [{\citenamefont {Komsa}\ \emph {et~al.}(2012)\citenamefont {Komsa},
  \citenamefont {Kotakoski}, \citenamefont {Kurasch}, \citenamefont {Lehtinen},
  \citenamefont {Kaiser},\ and\ \citenamefont {Krasheninnikov}}]{PRLkomsa2012}%
  \BibitemOpen
  \bibfield  {author} {\bibinfo {author} {\bibfnamefont {Hannu-Pekka}\
  \bibnamefont {Komsa}}, \bibinfo {author} {\bibfnamefont {Jani}\ \bibnamefont
  {Kotakoski}}, \bibinfo {author} {\bibfnamefont {Simon}\ \bibnamefont
  {Kurasch}}, \bibinfo {author} {\bibfnamefont {Ossi}\ \bibnamefont
  {Lehtinen}}, \bibinfo {author} {\bibfnamefont {Ute}\ \bibnamefont {Kaiser}},
  \ and\ \bibinfo {author} {\bibfnamefont {Arkady~V.}\ \bibnamefont
  {Krasheninnikov}},\ }\bibfield  {title} {\enquote {\bibinfo {title}
  {{Two-Dimensional Transition Metal Dichalcogenides under Electron
  Irradiation: Defect Production and Doping}},}\ }\href {\doibase
  10.1103/PhysRevLett.109.035503} {\bibfield  {journal} {\bibinfo  {journal}
  {Phys. Rev. Lett.}\ }\textbf {\bibinfo {volume} {109}},\ \bibinfo {pages}
  {035503} (\bibinfo {year} {2012})}\BibitemShut {NoStop}%
\bibitem [{\citenamefont {Guo}\ \emph {et~al.}(2021)\citenamefont {Guo},
  \citenamefont {Lin}, \citenamefont {Xie}, \citenamefont {Yuan}, \citenamefont
  {Zhu}, \citenamefont {Shen}, \citenamefont {Lu}, \citenamefont {Su},
  \citenamefont {Shi}, \citenamefont {Zhang}, \citenamefont {HuangFu},
  \citenamefont {Xu}, \citenamefont {Cai}, \citenamefont {Park}, \citenamefont
  {Ji}, \citenamefont {Wang}, \citenamefont {Dai}, \citenamefont {Tian},
  \citenamefont {Huang}, \citenamefont {Dou}, \citenamefont {Jiao},
  \citenamefont {Li}, \citenamefont {Yu}, \citenamefont {Idrobo}, \citenamefont
  {Cao}, \citenamefont {Palacios},\ and\ \citenamefont
  {Kong}}]{PNASyunfan2021}%
  \BibitemOpen
  \bibfield  {author} {\bibinfo {author} {\bibfnamefont {Yunfan}\ \bibnamefont
  {Guo}}, \bibinfo {author} {\bibfnamefont {Yuxuan}\ \bibnamefont {Lin}},
  \bibinfo {author} {\bibfnamefont {Kaichen}\ \bibnamefont {Xie}}, \bibinfo
  {author} {\bibfnamefont {Biao}\ \bibnamefont {Yuan}}, \bibinfo {author}
  {\bibfnamefont {Jiadi}\ \bibnamefont {Zhu}}, \bibinfo {author} {\bibfnamefont
  {Pin-Chun}\ \bibnamefont {Shen}}, \bibinfo {author} {\bibfnamefont {Ang-Yu}\
  \bibnamefont {Lu}}, \bibinfo {author} {\bibfnamefont {Cong}\ \bibnamefont
  {Su}}, \bibinfo {author} {\bibfnamefont {Enzheng}\ \bibnamefont {Shi}},
  \bibinfo {author} {\bibfnamefont {Kunyan}\ \bibnamefont {Zhang}}, \bibinfo
  {author} {\bibfnamefont {Changan}\ \bibnamefont {HuangFu}}, \bibinfo {author}
  {\bibfnamefont {Haowei}\ \bibnamefont {Xu}}, \bibinfo {author} {\bibfnamefont
  {Zhengyang}\ \bibnamefont {Cai}}, \bibinfo {author} {\bibfnamefont {Ji-Hoon}\
  \bibnamefont {Park}}, \bibinfo {author} {\bibfnamefont {Qingqing}\
  \bibnamefont {Ji}}, \bibinfo {author} {\bibfnamefont {Jiangtao}\ \bibnamefont
  {Wang}}, \bibinfo {author} {\bibfnamefont {Xiaochuan}\ \bibnamefont {Dai}},
  \bibinfo {author} {\bibfnamefont {Xuezeng}\ \bibnamefont {Tian}}, \bibinfo
  {author} {\bibfnamefont {Shengxi}\ \bibnamefont {Huang}}, \bibinfo {author}
  {\bibfnamefont {Letian}\ \bibnamefont {Dou}}, \bibinfo {author}
  {\bibfnamefont {Liying}\ \bibnamefont {Jiao}}, \bibinfo {author}
  {\bibfnamefont {Ju}~\bibnamefont {Li}}, \bibinfo {author} {\bibfnamefont
  {Yi}~\bibnamefont {Yu}}, \bibinfo {author} {\bibfnamefont {Juan-Carlos}\
  \bibnamefont {Idrobo}}, \bibinfo {author} {\bibfnamefont {Ting}\ \bibnamefont
  {Cao}}, \bibinfo {author} {\bibfnamefont {Tomás}\ \bibnamefont {Palacios}},
  \ and\ \bibinfo {author} {\bibfnamefont {Jing}\ \bibnamefont {Kong}},\
  }\bibfield  {title} {\enquote {\bibinfo {title} {{Designing artificial
  two-dimensional landscapes via atomic-layer substitution}},}\ }\href
  {\doibase 10.1073/pnas.2106124118} {\bibfield  {journal} {\bibinfo  {journal}
  {Proceedings of the National Academy of Sciences}\ }\textbf {\bibinfo
  {volume} {118}},\ \bibinfo {pages} {e2106124118} (\bibinfo {year}
  {2021})}\BibitemShut {NoStop}%
\bibitem [{\citenamefont {Qin}\ \emph {et~al.}(2022)\citenamefont {Qin},
  \citenamefont {Sayyad}, \citenamefont {Montblanch}, \citenamefont {Feuer},
  \citenamefont {Dey}, \citenamefont {Blei}, \citenamefont {Sailus},
  \citenamefont {Kara}, \citenamefont {Shen}, \citenamefont {Yang},
  \citenamefont {Botana}, \citenamefont {Atature},\ and\ \citenamefont
  {Tongay}}]{AMqin2022}%
  \BibitemOpen
  \bibfield  {author} {\bibinfo {author} {\bibfnamefont {Ying}\ \bibnamefont
  {Qin}}, \bibinfo {author} {\bibfnamefont {Mohammed}\ \bibnamefont {Sayyad}},
  \bibinfo {author} {\bibfnamefont {Alejandro R.-P.}\ \bibnamefont
  {Montblanch}}, \bibinfo {author} {\bibfnamefont {Matthew S.~G.}\ \bibnamefont
  {Feuer}}, \bibinfo {author} {\bibfnamefont {Dibyendu}\ \bibnamefont {Dey}},
  \bibinfo {author} {\bibfnamefont {Mark}\ \bibnamefont {Blei}}, \bibinfo
  {author} {\bibfnamefont {Renee}\ \bibnamefont {Sailus}}, \bibinfo {author}
  {\bibfnamefont {Dhiren~M.}\ \bibnamefont {Kara}}, \bibinfo {author}
  {\bibfnamefont {Yuxia}\ \bibnamefont {Shen}}, \bibinfo {author}
  {\bibfnamefont {Shize}\ \bibnamefont {Yang}}, \bibinfo {author}
  {\bibfnamefont {Antia~S.}\ \bibnamefont {Botana}}, \bibinfo {author}
  {\bibfnamefont {Mete}\ \bibnamefont {Atature}}, \ and\ \bibinfo {author}
  {\bibfnamefont {Sefaattin}\ \bibnamefont {Tongay}},\ }\bibfield  {title}
  {\enquote {\bibinfo {title} {{Reaching the Excitonic Limit in 2D Janus
  Monolayers by In Situ Deterministic Growth}},}\ }\href {\doibase
  https://doi.org/10.1002/adma.202106222} {\bibfield  {journal} {\bibinfo
  {journal} {Advanced Materials}\ }\textbf {\bibinfo {volume} {34}},\ \bibinfo
  {pages} {2106222} (\bibinfo {year} {2022})}\BibitemShut {NoStop}%
\bibitem [{\citenamefont {Das}\ \emph {et~al.}(2021)\citenamefont {Das},
  \citenamefont {Yang}, \citenamefont {Seo}, \citenamefont {Kim}, \citenamefont
  {Park}, \citenamefont {Kim}, \citenamefont {Seo}, \citenamefont {Kwak},\ and\
  \citenamefont {Chang}}]{ACSAMIdas2021}%
  \BibitemOpen
  \bibfield  {author} {\bibinfo {author} {\bibfnamefont {Tanmoy}\ \bibnamefont
  {Das}}, \bibinfo {author} {\bibfnamefont {Eunyeong}\ \bibnamefont {Yang}},
  \bibinfo {author} {\bibfnamefont {Jae~Eun}\ \bibnamefont {Seo}}, \bibinfo
  {author} {\bibfnamefont {Jeong~Hyeon}\ \bibnamefont {Kim}}, \bibinfo {author}
  {\bibfnamefont {Eunpyo}\ \bibnamefont {Park}}, \bibinfo {author}
  {\bibfnamefont {Minkyung}\ \bibnamefont {Kim}}, \bibinfo {author}
  {\bibfnamefont {Dongwook}\ \bibnamefont {Seo}}, \bibinfo {author}
  {\bibfnamefont {Joon~Young}\ \bibnamefont {Kwak}}, \ and\ \bibinfo {author}
  {\bibfnamefont {Jiwon}\ \bibnamefont {Chang}},\ }\bibfield  {title} {\enquote
  {\bibinfo {title} {{Doping-Free All PtSe2 Transistor via Thickness-Modulated
  Phase Transition}},}\ }\href {\doibase 10.1021/acsami.0c17810} {\bibfield
  {journal} {\bibinfo  {journal} {ACS Applied Materials \& Interfaces}\
  }\textbf {\bibinfo {volume} {13}},\ \bibinfo {pages} {1861--1871} (\bibinfo
  {year} {2021})}\BibitemShut {NoStop}%
\bibitem [{\citenamefont {Lin}\ \emph {et~al.}(2016)\citenamefont {Lin},
  \citenamefont {Carvalho}, \citenamefont {Kahn}, \citenamefont {Lv},
  \citenamefont {Rao}, \citenamefont {Terrones}, \citenamefont {Pimenta},\ and\
  \citenamefont {Terrones}}]{2DMlin2016}%
  \BibitemOpen
  \bibfield  {author} {\bibinfo {author} {\bibfnamefont {Zhong}\ \bibnamefont
  {Lin}}, \bibinfo {author} {\bibfnamefont {Bruno~R}\ \bibnamefont {Carvalho}},
  \bibinfo {author} {\bibfnamefont {Ethan}\ \bibnamefont {Kahn}}, \bibinfo
  {author} {\bibfnamefont {Ruitao}\ \bibnamefont {Lv}}, \bibinfo {author}
  {\bibfnamefont {Rahul}\ \bibnamefont {Rao}}, \bibinfo {author} {\bibfnamefont
  {Humberto}\ \bibnamefont {Terrones}}, \bibinfo {author} {\bibfnamefont
  {Marcos~A}\ \bibnamefont {Pimenta}}, \ and\ \bibinfo {author} {\bibfnamefont
  {Mauricio}\ \bibnamefont {Terrones}},\ }\bibfield  {title} {\enquote
  {\bibinfo {title} {{Defect engineering of two-dimensional transition metal
  dichalcogenides}},}\ }\href {\doibase 10.1088/2053-1583/3/2/022002}
  {\bibfield  {journal} {\bibinfo  {journal} {2D Materials}\ }\textbf {\bibinfo
  {volume} {3}},\ \bibinfo {pages} {022002} (\bibinfo {year}
  {2016})}\BibitemShut {NoStop}%
\bibitem [{\citenamefont {Focassio}\ \emph {et~al.}(2021)\citenamefont
  {Focassio}, \citenamefont {Schleder}, \citenamefont {Crasto~de Lima},
  \citenamefont {Lewenkopf},\ and\ \citenamefont {Fazzio}}]{PRBfocassio2021}%
  \BibitemOpen
  \bibfield  {author} {\bibinfo {author} {\bibfnamefont {Bruno}\ \bibnamefont
  {Focassio}}, \bibinfo {author} {\bibfnamefont {Gabriel~R.}\ \bibnamefont
  {Schleder}}, \bibinfo {author} {\bibfnamefont {Felipe}\ \bibnamefont
  {Crasto~de Lima}}, \bibinfo {author} {\bibfnamefont {Caio}\ \bibnamefont
  {Lewenkopf}}, \ and\ \bibinfo {author} {\bibfnamefont {Adalberto}\
  \bibnamefont {Fazzio}},\ }\bibfield  {title} {\enquote {\bibinfo {title}
  {{Amorphous ${\mathrm{Bi}}_{2}{\mathrm{Se}}_{3}$ structural, electronic, and
  topological nature from first principles}},}\ }\href {\doibase
  10.1103/PhysRevB.104.214206} {\bibfield  {journal} {\bibinfo  {journal}
  {Phys. Rev. B}\ }\textbf {\bibinfo {volume} {104}},\ \bibinfo {pages}
  {214206} (\bibinfo {year} {2021})}\BibitemShut {NoStop}%
\bibitem [{\citenamefont {Qiu}\ \emph {et~al.}(2013)\citenamefont {Qiu},
  \citenamefont {Xu}, \citenamefont {Wang}, \citenamefont {Ren}, \citenamefont
  {Nan}, \citenamefont {Ni}, \citenamefont {Chen}, \citenamefont {Yuan},
  \citenamefont {Miao}, \citenamefont {Song}, \citenamefont {Long},
  \citenamefont {Shi}, \citenamefont {Sun}, \citenamefont {Wang},\ and\
  \citenamefont {Wang}}]{NATCOMMqiu2013}%
  \BibitemOpen
  \bibfield  {author} {\bibinfo {author} {\bibfnamefont {Hao}\ \bibnamefont
  {Qiu}}, \bibinfo {author} {\bibfnamefont {Tao}\ \bibnamefont {Xu}}, \bibinfo
  {author} {\bibfnamefont {Zilu}\ \bibnamefont {Wang}}, \bibinfo {author}
  {\bibfnamefont {Wei}\ \bibnamefont {Ren}}, \bibinfo {author} {\bibfnamefont
  {Haiyan}\ \bibnamefont {Nan}}, \bibinfo {author} {\bibfnamefont {Zhenhua}\
  \bibnamefont {Ni}}, \bibinfo {author} {\bibfnamefont {Qian}\ \bibnamefont
  {Chen}}, \bibinfo {author} {\bibfnamefont {Shijun}\ \bibnamefont {Yuan}},
  \bibinfo {author} {\bibfnamefont {Feng}\ \bibnamefont {Miao}}, \bibinfo
  {author} {\bibfnamefont {Fengqi}\ \bibnamefont {Song}}, \bibinfo {author}
  {\bibfnamefont {Gen}\ \bibnamefont {Long}}, \bibinfo {author} {\bibfnamefont
  {Yi}~\bibnamefont {Shi}}, \bibinfo {author} {\bibfnamefont {Litao}\
  \bibnamefont {Sun}}, \bibinfo {author} {\bibfnamefont {Jinlan}\ \bibnamefont
  {Wang}}, \ and\ \bibinfo {author} {\bibfnamefont {Xinran}\ \bibnamefont
  {Wang}},\ }\bibfield  {title} {\enquote {\bibinfo {title} {{Hopping transport
  through defect-induced localized states in molybdenum disulphide}},}\ }\href
  {\doibase 10.1038/ncomms3642} {\bibfield  {journal} {\bibinfo  {journal}
  {Nature Communications}\ }\textbf {\bibinfo {volume} {4}},\ \bibinfo {pages}
  {2642} (\bibinfo {year} {2013})}\BibitemShut {NoStop}%
\bibitem [{\citenamefont {Crasto~de Lima}\ and\ \citenamefont
  {Fazzio}(2021)}]{NLcrasto2021}%
  \BibitemOpen
  \bibfield  {author} {\bibinfo {author} {\bibfnamefont {Felipe}\ \bibnamefont
  {Crasto~de Lima}}\ and\ \bibinfo {author} {\bibfnamefont {Adalberto}\
  \bibnamefont {Fazzio}},\ }\bibfield  {title} {\enquote {\bibinfo {title} {{At
  the Verge of Topology: Vacancy-Driven Quantum Spin Hall in Trivial
  Insulators}},}\ }\href {\doibase 10.1021/acs.nanolett.1c02458} {\bibfield
  {journal} {\bibinfo  {journal} {Nano Letters}\ }\textbf {\bibinfo {volume}
  {21}},\ \bibinfo {pages} {9398--9402} (\bibinfo {year} {2021})}\BibitemShut
  {NoStop}%
\bibitem [{\citenamefont {Zhao}\ \emph {et~al.}(2014)\citenamefont {Zhao},
  \citenamefont {Zhang}, \citenamefont {Feng}, \citenamefont {Yao},\ and\
  \citenamefont {Yang}}]{PRBzhao2014}%
  \BibitemOpen
  \bibfield  {author} {\bibinfo {author} {\bibfnamefont {Bao}\ \bibnamefont
  {Zhao}}, \bibinfo {author} {\bibfnamefont {Jiayong}\ \bibnamefont {Zhang}},
  \bibinfo {author} {\bibfnamefont {Wanxiang}\ \bibnamefont {Feng}}, \bibinfo
  {author} {\bibfnamefont {Yugui}\ \bibnamefont {Yao}}, \ and\ \bibinfo
  {author} {\bibfnamefont {Zhongqin}\ \bibnamefont {Yang}},\ }\bibfield
  {title} {\enquote {\bibinfo {title} {{Quantum spin Hall and Z$_2$ metallic
  states in an organic material}},}\ }\href {\doibase
  10.1103/PhysRevB.90.201403} {\bibfield  {journal} {\bibinfo  {journal} {Phys.
  Rev. B}\ }\textbf {\bibinfo {volume} {90}},\ \bibinfo {pages} {201403}
  (\bibinfo {year} {2014})}\BibitemShut {NoStop}%
\bibitem [{\citenamefont {Kresse}\ and\ \citenamefont
  {Furthm\"uller}(1996)}]{vasp1}%
  \BibitemOpen
  \bibfield  {author} {\bibinfo {author} {\bibfnamefont {G.}~\bibnamefont
  {Kresse}}\ and\ \bibinfo {author} {\bibfnamefont {J.}~\bibnamefont
  {Furthm\"uller}},\ }\bibfield  {title} {\enquote {\bibinfo {title}
  {{Efficiency of ab-initio total energy calculations for metals and
  semiconductors using a plane-wave basis set}},}\ }\href {\doibase
  https://doi.org/10.1016/0927-0256(96)00008-0} {\bibfield  {journal} {\bibinfo
   {journal} {Computational Materials Science}\ }\textbf {\bibinfo {volume}
  {6}},\ \bibinfo {pages} {15 -- 50} (\bibinfo {year} {1996})}\BibitemShut
  {NoStop}%
\bibitem [{\citenamefont {Perdew}\ \emph {et~al.}(1996)\citenamefont {Perdew},
  \citenamefont {Burke},\ and\ \citenamefont {Ernzerhof}}]{PBE}%
  \BibitemOpen
  \bibfield  {author} {\bibinfo {author} {\bibfnamefont {J.~P.}\ \bibnamefont
  {Perdew}}, \bibinfo {author} {\bibfnamefont {K.}~\bibnamefont {Burke}}, \
  and\ \bibinfo {author} {\bibfnamefont {M.}~\bibnamefont {Ernzerhof}},\
  }\bibfield  {title} {\enquote {\bibinfo {title} {{Generalized Gradient
  Approximation Made Simple}},}\ }\href@noop {} {\bibfield  {journal} {\bibinfo
   {journal} {Phys. Rev. Lett.}\ }\textbf {\bibinfo {volume} {77}},\ \bibinfo
  {pages} {3865} (\bibinfo {year} {1996})}\BibitemShut {NoStop}%
\bibitem [{\citenamefont {Bl\"ochl}(1994)}]{PRBblochl1994}%
  \BibitemOpen
  \bibfield  {author} {\bibinfo {author} {\bibfnamefont {P.~E.}\ \bibnamefont
  {Bl\"ochl}},\ }\bibfield  {title} {\enquote {\bibinfo {title} {{Projector
  augmented-wave method}},}\ }\href {\doibase 10.1103/PhysRevB.50.17953}
  {\bibfield  {journal} {\bibinfo  {journal} {Phys. Rev. B}\ }\textbf {\bibinfo
  {volume} {50}},\ \bibinfo {pages} {17953--17979} (\bibinfo {year}
  {1994})}\BibitemShut {NoStop}%
\bibitem [{\citenamefont {Soluyanov}\ and\ \citenamefont
  {Vanderbilt}(2011)}]{PRBsoluyanov2011}%
  \BibitemOpen
  \bibfield  {author} {\bibinfo {author} {\bibfnamefont {Alexey~A.}\
  \bibnamefont {Soluyanov}}\ and\ \bibinfo {author} {\bibfnamefont {David}\
  \bibnamefont {Vanderbilt}},\ }\bibfield  {title} {\enquote {\bibinfo {title}
  {{Computing topological invariants without inversion symmetry}},}\ }\href
  {\doibase 10.1103/PhysRevB.83.235401} {\bibfield  {journal} {\bibinfo
  {journal} {Phys. Rev. B}\ }\textbf {\bibinfo {volume} {83}},\ \bibinfo
  {pages} {235401} (\bibinfo {year} {2011})}\BibitemShut {NoStop}%
\bibitem [{\citenamefont {Yu}\ \emph {et~al.}(2011)\citenamefont {Yu},
  \citenamefont {Qi}, \citenamefont {Bernevig}, \citenamefont {Fang},\ and\
  \citenamefont {Dai}}]{PRByu2011}%
  \BibitemOpen
  \bibfield  {author} {\bibinfo {author} {\bibfnamefont {Rui}\ \bibnamefont
  {Yu}}, \bibinfo {author} {\bibfnamefont {Xiao~Liang}\ \bibnamefont {Qi}},
  \bibinfo {author} {\bibfnamefont {Andrei}\ \bibnamefont {Bernevig}}, \bibinfo
  {author} {\bibfnamefont {Zhong}\ \bibnamefont {Fang}}, \ and\ \bibinfo
  {author} {\bibfnamefont {Xi}~\bibnamefont {Dai}},\ }\bibfield  {title}
  {\enquote {\bibinfo {title} {{Equivalent expression of ${\mathbb{Z}}_{2}$
  topological invariant for band insulators using the non-Abelian Berry
  connection}},}\ }\href {\doibase 10.1103/PhysRevB.84.075119} {\bibfield
  {journal} {\bibinfo  {journal} {Phys. Rev. B}\ }\textbf {\bibinfo {volume}
  {84}},\ \bibinfo {pages} {075119} (\bibinfo {year} {2011})}\BibitemShut
  {NoStop}%
\bibitem [{\citenamefont {Huang}\ and\ \citenamefont
  {Liu}(2018)}]{PRBhuang2018}%
  \BibitemOpen
  \bibfield  {author} {\bibinfo {author} {\bibfnamefont {Huaqing}\ \bibnamefont
  {Huang}}\ and\ \bibinfo {author} {\bibfnamefont {Feng}\ \bibnamefont {Liu}},\
  }\bibfield  {title} {\enquote {\bibinfo {title} {{Theory of spin Bott index
  for quantum spin Hall states in nonperiodic systems}},}\ }\href {\doibase
  10.1103/PhysRevB.98.125130} {\bibfield  {journal} {\bibinfo  {journal} {Phys.
  Rev. B}\ }\textbf {\bibinfo {volume} {98}},\ \bibinfo {pages} {125130}
  (\bibinfo {year} {2018})}\BibitemShut {NoStop}%
\bibitem [{\citenamefont {Mostofi}\ \emph {et~al.}(2014)\citenamefont
  {Mostofi}, \citenamefont {Yates}, \citenamefont {Pizzi}, \citenamefont {Lee},
  \citenamefont {Souza}, \citenamefont {Vanderbilt},\ and\ \citenamefont
  {Marzari}}]{CPCmostofi2014}%
  \BibitemOpen
  \bibfield  {author} {\bibinfo {author} {\bibfnamefont {Arash~A.}\
  \bibnamefont {Mostofi}}, \bibinfo {author} {\bibfnamefont {Jonathan~R.}\
  \bibnamefont {Yates}}, \bibinfo {author} {\bibfnamefont {Giovanni}\
  \bibnamefont {Pizzi}}, \bibinfo {author} {\bibfnamefont {Young-Su}\
  \bibnamefont {Lee}}, \bibinfo {author} {\bibfnamefont {Ivo}\ \bibnamefont
  {Souza}}, \bibinfo {author} {\bibfnamefont {David}\ \bibnamefont
  {Vanderbilt}}, \ and\ \bibinfo {author} {\bibfnamefont {Nicola}\ \bibnamefont
  {Marzari}},\ }\bibfield  {title} {\enquote {\bibinfo {title} {{An updated
  version of wannier90: A tool for obtaining maximally-localised Wannier
  functions}},}\ }\href {\doibase https://doi.org/10.1016/j.cpc.2014.05.003}
  {\bibfield  {journal} {\bibinfo  {journal} {Computer Physics Communications}\
  }\textbf {\bibinfo {volume} {185}},\ \bibinfo {pages} {2309 -- 2310}
  (\bibinfo {year} {2014})}\BibitemShut {NoStop}%
\bibitem [{\citenamefont {Zunger}\ \emph {et~al.}(1990)\citenamefont {Zunger},
  \citenamefont {Wei}, \citenamefont {Ferreira},\ and\ \citenamefont
  {Bernard}}]{PRLzunger1990}%
  \BibitemOpen
  \bibfield  {author} {\bibinfo {author} {\bibfnamefont {Alex}\ \bibnamefont
  {Zunger}}, \bibinfo {author} {\bibfnamefont {S.-H.}\ \bibnamefont {Wei}},
  \bibinfo {author} {\bibfnamefont {L.~G.}\ \bibnamefont {Ferreira}}, \ and\
  \bibinfo {author} {\bibfnamefont {James~E.}\ \bibnamefont {Bernard}},\
  }\bibfield  {title} {\enquote {\bibinfo {title} {{Special quasirandom
  structures}},}\ }\href {\doibase 10.1103/PhysRevLett.65.353} {\bibfield
  {journal} {\bibinfo  {journal} {Phys. Rev. Lett.}\ }\textbf {\bibinfo
  {volume} {65}},\ \bibinfo {pages} {353--356} (\bibinfo {year}
  {1990})}\BibitemShut {NoStop}%
\bibitem [{\citenamefont {Absor}\ \emph {et~al.}(2017)\citenamefont {Absor},
  \citenamefont {Santoso}, \citenamefont {Abraha}, \citenamefont {Ishii},
  \citenamefont {Saito} \emph {et~al.}}]{absor2017defect}%
  \BibitemOpen
  \bibfield  {author} {\bibinfo {author} {\bibfnamefont {Moh Adhib~Ulil}\
  \bibnamefont {Absor}}, \bibinfo {author} {\bibfnamefont {Iman}\ \bibnamefont
  {Santoso}}, \bibinfo {author} {\bibfnamefont {Kamsul}\ \bibnamefont
  {Abraha}}, \bibinfo {author} {\bibfnamefont {Fumiyuki}\ \bibnamefont
  {Ishii}}, \bibinfo {author} {\bibfnamefont {Mineo}\ \bibnamefont {Saito}},
  \emph {et~al.},\ }\bibfield  {title} {\enquote {\bibinfo {title}
  {{Defect-induced large spin-orbit splitting in monolayer PtSe2}},}\
  }\href@noop {} {\bibfield  {journal} {\bibinfo  {journal} {Physical Review
  B}\ }\textbf {\bibinfo {volume} {96}},\ \bibinfo {pages} {115128} (\bibinfo
  {year} {2017})}\BibitemShut {NoStop}%
\bibitem [{\citenamefont {Freire}\ \emph {et~al.}(2022)\citenamefont {Freire},
  \citenamefont {de~Lima},\ and\ \citenamefont {Fazzio}}]{freire2022vacancy}%
  \BibitemOpen
  \bibfield  {author} {\bibinfo {author} {\bibfnamefont {Rafael~LH}\
  \bibnamefont {Freire}}, \bibinfo {author} {\bibfnamefont {Felipe~Crasto}\
  \bibnamefont {de~Lima}}, \ and\ \bibinfo {author} {\bibfnamefont {Adalberto}\
  \bibnamefont {Fazzio}},\ }\bibfield  {title} {\enquote {\bibinfo {title}
  {{Vacancy localization effects on MX2 transition metal dichalcogenides: a
  systematic ab-initio study}},}\ }\href@noop {} {\bibfield  {journal}
  {\bibinfo  {journal} {Physical Review Materials}\ }\textbf {\bibinfo {volume}
  {6}},\ \bibinfo {pages} {084002} (\bibinfo {year} {2022})}\BibitemShut
  {NoStop}%
\bibitem [{\citenamefont {Lin}\ \emph {et~al.}(2017)\citenamefont {Lin},
  \citenamefont {Lu}, \citenamefont {Shao}, \citenamefont {Zhang},
  \citenamefont {Wu}, \citenamefont {Pan}, \citenamefont {Gao}, \citenamefont
  {Zhu}, \citenamefont {Qian}, \citenamefont {Zhang}, \citenamefont {Bao},
  \citenamefont {Li}, \citenamefont {Wang}, \citenamefont {Liu}, \citenamefont
  {Sun}, \citenamefont {Lei}, \citenamefont {Liu}, \citenamefont {Wang},
  \citenamefont {Ibrahim}, \citenamefont {Leonard}, \citenamefont {Zhou},
  \citenamefont {Guo}, \citenamefont {Wang}, \citenamefont {Du}, \citenamefont
  {Pantelides},\ and\ \citenamefont {Gao}}]{NATMATlin2017}%
  \BibitemOpen
  \bibfield  {author} {\bibinfo {author} {\bibfnamefont {X.}~\bibnamefont
  {Lin}}, \bibinfo {author} {\bibfnamefont {J.~C.}\ \bibnamefont {Lu}},
  \bibinfo {author} {\bibfnamefont {Y.}~\bibnamefont {Shao}}, \bibinfo {author}
  {\bibfnamefont {Y.~Y.}\ \bibnamefont {Zhang}}, \bibinfo {author}
  {\bibfnamefont {X.}~\bibnamefont {Wu}}, \bibinfo {author} {\bibfnamefont
  {J.~B.}\ \bibnamefont {Pan}}, \bibinfo {author} {\bibfnamefont
  {L.}~\bibnamefont {Gao}}, \bibinfo {author} {\bibfnamefont {S.~Y.}\
  \bibnamefont {Zhu}}, \bibinfo {author} {\bibfnamefont {K.}~\bibnamefont
  {Qian}}, \bibinfo {author} {\bibfnamefont {Y.~F.}\ \bibnamefont {Zhang}},
  \bibinfo {author} {\bibfnamefont {D.~L.}\ \bibnamefont {Bao}}, \bibinfo
  {author} {\bibfnamefont {L.~F.}\ \bibnamefont {Li}}, \bibinfo {author}
  {\bibfnamefont {Y.~Q.}\ \bibnamefont {Wang}}, \bibinfo {author}
  {\bibfnamefont {Z.~L.}\ \bibnamefont {Liu}}, \bibinfo {author} {\bibfnamefont
  {J.~T.}\ \bibnamefont {Sun}}, \bibinfo {author} {\bibfnamefont
  {T.}~\bibnamefont {Lei}}, \bibinfo {author} {\bibfnamefont {C.}~\bibnamefont
  {Liu}}, \bibinfo {author} {\bibfnamefont {J.~O.}\ \bibnamefont {Wang}},
  \bibinfo {author} {\bibfnamefont {K.}~\bibnamefont {Ibrahim}}, \bibinfo
  {author} {\bibfnamefont {D.~N.}\ \bibnamefont {Leonard}}, \bibinfo {author}
  {\bibfnamefont {W.}~\bibnamefont {Zhou}}, \bibinfo {author} {\bibfnamefont
  {H.~M.}\ \bibnamefont {Guo}}, \bibinfo {author} {\bibfnamefont {Y.~L.}\
  \bibnamefont {Wang}}, \bibinfo {author} {\bibfnamefont {S.~X.}\ \bibnamefont
  {Du}}, \bibinfo {author} {\bibfnamefont {S.~T.}\ \bibnamefont {Pantelides}},
  \ and\ \bibinfo {author} {\bibfnamefont {H.-J.}\ \bibnamefont {Gao}},\
  }\bibfield  {title} {\enquote {\bibinfo {title} {{Intrinsically patterned
  two-dimensional materials for selective adsorption of molecules and
  nanoclusters}},}\ }\href {\doibase 10.1038/nmat4915} {\bibfield  {journal}
  {\bibinfo  {journal} {Nature Materials}\ }\textbf {\bibinfo {volume} {16}},\
  \bibinfo {pages} {717--721} (\bibinfo {year} {2017})}\BibitemShut {NoStop}%
\bibitem [{\citenamefont {Chen}\ \emph {et~al.}(2022)\citenamefont {Chen},
  \citenamefont {Zhou}, \citenamefont {Xu}, \citenamefont {Wen}, \citenamefont
  {Liu},\ and\ \citenamefont {Warner}}]{NLchen2022}%
  \BibitemOpen
  \bibfield  {author} {\bibinfo {author} {\bibfnamefont {Jun}\ \bibnamefont
  {Chen}}, \bibinfo {author} {\bibfnamefont {Jiang}\ \bibnamefont {Zhou}},
  \bibinfo {author} {\bibfnamefont {Wenshuo}\ \bibnamefont {Xu}}, \bibinfo
  {author} {\bibfnamefont {Yi}~\bibnamefont {Wen}}, \bibinfo {author}
  {\bibfnamefont {Yuanyue}\ \bibnamefont {Liu}}, \ and\ \bibinfo {author}
  {\bibfnamefont {Jamie~H.}\ \bibnamefont {Warner}},\ }\bibfield  {title}
  {\enquote {\bibinfo {title} {{Atomic-Level Dynamics of Point Vacancies and
  the Induced Stretched Defects in 2D Monolayer PtSe2}},}\ }\href {\doibase
  10.1021/acs.nanolett.1c04275} {\bibfield  {journal} {\bibinfo  {journal}
  {Nano Letters}\ }\textbf {\bibinfo {volume} {22}},\ \bibinfo {pages}
  {3289--3297} (\bibinfo {year} {2022})}\BibitemShut {NoStop}%
\bibitem [{\citenamefont {Mitchell}\ \emph {et~al.}(2018)\citenamefont
  {Mitchell}, \citenamefont {Nash}, \citenamefont {Hexner}, \citenamefont
  {Turner},\ and\ \citenamefont {Irvine}}]{NATPHYSmitchell2018}%
  \BibitemOpen
  \bibfield  {author} {\bibinfo {author} {\bibfnamefont {Noah~P.}\ \bibnamefont
  {Mitchell}}, \bibinfo {author} {\bibfnamefont {Lisa~M.}\ \bibnamefont
  {Nash}}, \bibinfo {author} {\bibfnamefont {Daniel}\ \bibnamefont {Hexner}},
  \bibinfo {author} {\bibfnamefont {Ari~M.}\ \bibnamefont {Turner}}, \ and\
  \bibinfo {author} {\bibfnamefont {William T.~M.}\ \bibnamefont {Irvine}},\
  }\bibfield  {title} {\enquote {\bibinfo {title} {{Amorphous topological
  insulators constructed from random point sets}},}\ }\href {\doibase
  10.1038/s41567-017-0024-5} {\bibfield  {journal} {\bibinfo  {journal} {Nature
  Physics}\ }\textbf {\bibinfo {volume} {14}},\ \bibinfo {pages} {380--385}
  (\bibinfo {year} {2018})}\BibitemShut {NoStop}%
\bibitem [{\citenamefont {Agarwala}\ and\ \citenamefont
  {Shenoy}(2017)}]{PRLagarwala2017}%
  \BibitemOpen
  \bibfield  {author} {\bibinfo {author} {\bibfnamefont {Adhip}\ \bibnamefont
  {Agarwala}}\ and\ \bibinfo {author} {\bibfnamefont {Vijay~B.}\ \bibnamefont
  {Shenoy}},\ }\bibfield  {title} {\enquote {\bibinfo {title} {{Topological
  Insulators in Amorphous Systems}},}\ }\href {\doibase
  10.1103/PhysRevLett.118.236402} {\bibfield  {journal} {\bibinfo  {journal}
  {Phys. Rev. Lett.}\ }\textbf {\bibinfo {volume} {118}},\ \bibinfo {pages}
  {236402} (\bibinfo {year} {2017})}\BibitemShut {NoStop}%
\bibitem [{\citenamefont {Costa}\ \emph {et~al.}(2019)\citenamefont {Costa},
  \citenamefont {Schleder}, \citenamefont {Buongiorno~Nardelli}, \citenamefont
  {Lewenkopf},\ and\ \citenamefont {Fazzio}}]{NLcosta2019}%
  \BibitemOpen
  \bibfield  {author} {\bibinfo {author} {\bibfnamefont {Marcio}\ \bibnamefont
  {Costa}}, \bibinfo {author} {\bibfnamefont {Gabriel~R.}\ \bibnamefont
  {Schleder}}, \bibinfo {author} {\bibfnamefont {Marco}\ \bibnamefont
  {Buongiorno~Nardelli}}, \bibinfo {author} {\bibfnamefont {Caio}\ \bibnamefont
  {Lewenkopf}}, \ and\ \bibinfo {author} {\bibfnamefont {Adalberto}\
  \bibnamefont {Fazzio}},\ }\bibfield  {title} {\enquote {\bibinfo {title}
  {{Toward Realistic Amorphous Topological Insulators}},}\ }\href {\doibase
  10.1021/acs.nanolett.9b03881} {\bibfield  {journal} {\bibinfo  {journal}
  {Nano Letters}\ }\textbf {\bibinfo {volume} {19}},\ \bibinfo {pages}
  {8941--8946} (\bibinfo {year} {2019})}\BibitemShut {NoStop}%
\bibitem [{\citenamefont {Lee}\ \emph {et~al.}(2014)\citenamefont {Lee},
  \citenamefont {Huang},\ and\ \citenamefont {Mou}}]{JPCMlee2014}%
  \BibitemOpen
  \bibfield  {author} {\bibinfo {author} {\bibfnamefont {Shi-Ting}\
  \bibnamefont {Lee}}, \bibinfo {author} {\bibfnamefont {Shin-Ming}\
  \bibnamefont {Huang}}, \ and\ \bibinfo {author} {\bibfnamefont {Chung-Yu}\
  \bibnamefont {Mou}},\ }\bibfield  {title} {\enquote {\bibinfo {title}
  {{Stability of Z2 topological order in the presence of vacancy-induced
  impurity band}},}\ }\href {\doibase 10.1088/0953-8984/26/25/255502}
  {\bibfield  {journal} {\bibinfo  {journal} {Journal of Physics: Condensed
  Matter}\ }\textbf {\bibinfo {volume} {26}},\ \bibinfo {pages} {255502}
  (\bibinfo {year} {2014})}\BibitemShut {NoStop}%
\bibitem [{\citenamefont {Suding}\ and\ \citenamefont
  {Ziff}(1999)}]{PREsuding1999}%
  \BibitemOpen
  \bibfield  {author} {\bibinfo {author} {\bibfnamefont {Paul~N.}\ \bibnamefont
  {Suding}}\ and\ \bibinfo {author} {\bibfnamefont {Robert~M.}\ \bibnamefont
  {Ziff}},\ }\bibfield  {title} {\enquote {\bibinfo {title} {{Site percolation
  thresholds for Archimedean lattices}},}\ }\href {\doibase
  10.1103/PhysRevE.60.275} {\bibfield  {journal} {\bibinfo  {journal} {Phys.
  Rev. E}\ }\textbf {\bibinfo {volume} {60}},\ \bibinfo {pages} {275--283}
  (\bibinfo {year} {1999})}\BibitemShut {NoStop}%
\end{thebibliography}%

\end{document}